\documentclass{IEEEcsmag}

\usepackage[colorlinks,urlcolor=blue,linkcolor=blue,citecolor=blue]{hyperref}
\expandafter\def\expandafter\UrlBreaks\expandafter{\UrlBreaks\do\/\do\*\do\-\do\~\do\'\do\"\do\-}
\usepackage{upmath,color}

\usepackage{mathtools}

\let\labelindent\relax
\usepackage{enumitem}

\jvol{XX}
\jnum{XX}
\paper{8}
\jmonth{Month}
\jname{Software}
\jtitle{Design Patterns for Machine Learning Based
Systems with Human-in-the-Loop}
\pubyear{2023}

\begin{document}

\sptitle{Preprint - Authors' version - Accepted for publication at IEEE Software}

\title{Design Patterns for Machine Learning Based Systems with Human-in-the-Loop}

\author{Jakob Smedegaard Andersen and Walid Maalej}
\affil{University of Hamburg, Germany}

\markboth{THEME/FEATURE/DEPARTMENT}{THEME/FEATURE/DEPARTMENT}

\begin{abstract}\looseness-1
The development and deployment of systems using supervised machine learning (ML) remain challenging: mainly due to the limited reliability of prediction models and the lack of knowledge on how to effectively integrate human intelligence into automated decision-making. 
Humans involvement in the ML process is a promising and powerful paradigm to overcome the limitations of pure automated predictions and improve the applicability of ML in practice. 
We compile a catalog of design patterns to guide developers select and implement suitable human-in-the-loop (HiL) solutions.
Our catalog takes into consideration key requirements as the cost of human involvement and model retraining. 
It includes four training patterns, four deployment patterns, and two orthogonal cooperation patterns. 

\end{abstract}

\maketitle

\chapteri{M}achine learning (ML) has become a major field of research with tremendous progress in recent years.
With this, software developers are increasingly confronted with the need to integrate ML models into their systems. 
In particular, supervised ML \iffalse\cite{alpaydin2020introduction}\fi has become  essential  to leverage insights hidden in large datasets,  optimize workflows, and gain competitive advantages.
Supervised ML aims at learning patterns from pre-labelled examples in order to make accurate predictions about new unseen data.

Despite the research progress, the development and deployment of ML approaches for real world applications still poses critical engineering \cite{ishikawa2019engineers, Maalej:Compter:23}
and deployment challenges \cite{beede2020human}. 
Getting ML models to work accurately in practice is drastically different from idealized research and development environments.
Typical limitations include not meeting accuracy and reliability requirements, the lack of
user acceptance, model hallucination, few but critical prediction mistakes, as well as the shortage of adequate and rich  data to (re)train models.
To mitigate the limitations of pure automated approaches, research has thus suggested the Human-in-the-Loop (HiL) paradigm \cite{holzinger2016interactive}.

\begin{figure*}
    \centerline{\includegraphics[width=0.9\linewidth]{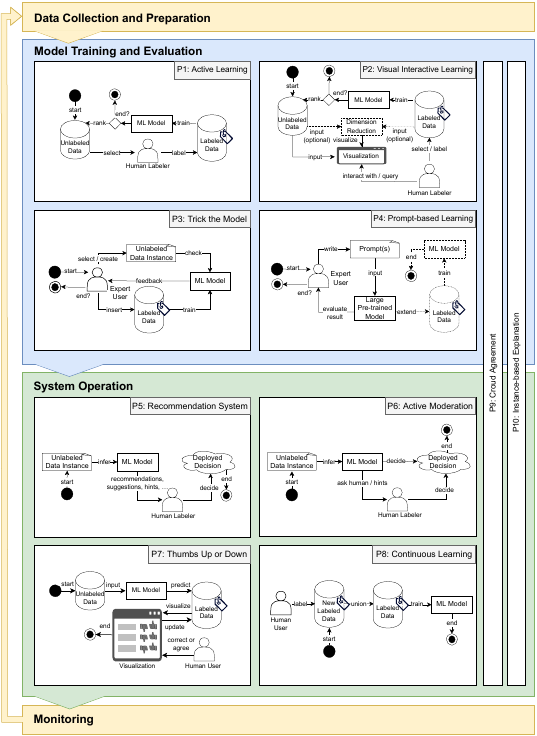}}
    \caption{HiL patterns along the ML pipeline.}
    \label{fig:hil_ptterns}
\end{figure*}

In HiL, humans are empowered to continuously provide \textit{feedback to the system} at each stage of the ML pipeline: including data collection and preparation, model training and evaluation, operation and monitoring (as Figure \ref{fig:hil_ptterns} shows). 
HiL focuses on feedback from domain experts and actual users of the system, i.e.~people with expertise in a specific application domain but not necessarily technical knowledge.
For supervised ML, this feedback is usually either  corrections of labels or additional labels. 
Domain-, application- or context-specific feedback has the advantage of incorporating specific knowledge from the actual domain, which a model is unaware of---thus with a high potential for improvement.
HiL considers human feedback an essential part of the ML process, rather than replacing humans with full automation. 
Simple, time-consuming, and repetitive tasks should be automated as much as possible, while the human should step into the loop from time to time to solve creative, challenging, difficult, or interesting cases.

While the concept behind HiL is promising, its research space is rather convoluted and its operationalizability and applicability in software development is rather unexplored.  
Moreover, the design space requires a careful  consideration: 
\begin{itemize}
    \item Human assistance in the creation and use of ML models usually comes with a high cost and should thus be carefully designed and minimized. 
    \item Also retraining ML models can, in certain cases, be very expensive and with significant carbon footprint. 
    \item Both models and humans may make (different types of) mistakes which require a careful trade-off discussion and decision that fits the use cases and domains at hand. 
\end{itemize}

To raise awareness about HiL and guide software developers systematically choose and adapt HiL solutions that fit their requirements, we propose a catalog of ten patterns for designing ML-based systems. 
While HiL has been around for a while as a paradigm to design and evaluate ML models, we focus on the software engineering perspective providing recurring, reusable HiL solutions for designing ML-based systems given certain contexts.

We define a “design pattern for HiL” as any pattern that directly or indirectly addresses the learning behavior or prediction outcome of an ML-based system via human feedback. 
Patterns are general solutions for common, recurring problems faced by software developers with best practices to implement them. 
Our catalog of patterns provides a shared vocabulary to better communicate the proven HiL design solutions. 
While software engineering patterns have been suggested for designing ML systems in general \cite{washizaki2022software}, there is a lack of  knowledge on how to effectively design and deploy the HiL paradigm.

Our catalog is the result of a preliminary semi-systematic literature review (including gray literature) combined with our own experience with researching and developing HiL systems \cite{settles1995active,bernard2018vial,attenberg2011beat,brown2020language,sengupta2017radar,andersen2022efficient,shukla2021agileml,karmakharm2019journalist}. %[3,5,6,8,11,14].
We queried Google and Google Scholar for papers and applications of HiL in the context of supervised ML, using the query string: '\textit{human-in-the-loop AND ("machine learning" OR classification OR regression)}'. We then checked the first 100 results for reusable solutions that focus on engaging users or domain experts, rather than, e.g., data scientists and engineers. 
Our proposed catalog reflect popular patterns with practical utility. 
It is likely incomplete and can be further extended.

\section{HiL Training Patterns}

Supervised ML models often require a large amount of labeled data to be properly trained. 
In the following, we outline design patterns for a cost-efficient and effective HiL training process of ML models.

% # # # # # # # # # # # # # # # # # 
% # A C T I V E   L E A R N I N G #
% # # # # # # # # # # # # # # # # # 

\subsection{P1: Active Learning \cite{settles1995active} } 

\textbf{Goal:} Minimize the number of labels needed to train a highly accurate ML model. \hfill\\
 \textbf{Human Role:} Label a sequence of automatically selected data instances.
 \hfill\\ \textbf{Problem:} ML models require numerous high-quality training examples to reliably recognize patterns in unseen data. 
 Since training examples are typically sparse or unavailable, they must be collected and labeled at hand. 
 This is a labor- and cost-intensive task that might be impractical on a large scale. It is thus desirable to minimize the human effort during the labeling of the initial training data. \hfill\\
\textbf{Structure:} 
In Active Learning, a model actively selects its own training data from a typically large corpora of unlabeled data instances. 
Carefully selecting highly informative and expressive instances can drastically reduce the amount of training data---and thus the cost of the labeling process---needed to achieve a given level of performance. 
The process of Active Learning consists of the following steps: 
First, potential unlabeled training instances are ranked according to their expected contribution to a model's learning behavior. Such information can be extracted from a model using uncertainty sampling \cite{settles1995active,andersen2022efficient}. The $k$-highest ranked instances %that have the highest uncertainty 
are then labeled by a human oracle and added to the training data. 
The model is then retrained until a  criterion is met, e.g., the labeling budget is depleted, or the model accuracy is acceptable. \hfill\\
\textbf{Impact:}  
Active Learning allows saving human resources and time during the development and (pre-) deployment. %Humans are involved by responding to a flow of label requests. 
\hfill\\
\textbf{Context:}
There is a lack of labeled examples, but a large amount of unlabeled data available.
An extensive user interface is not required.
\hfill\\
\textbf{Challenges:}
\begin{itemize}
    \item High dependency between training data and learned model.
    \item Requires low-latency training loops to maintain usability.
    \item Tendency to be boring and tedious at scale.
\end{itemize}
\textbf{Examples}: 
Active Anno\footnote{\url{https://github.com/MaxMello/ActiveAnno}},
Alanno\footnote{\url{https://github.com/josipjukic/alanno}}, and
Encord\footnote{\url{https://encord.com/}}

% # # # # # # # # # # # # # # # # # # # # # # # # # # # # # 
% # V I S U A L   I N T E R A C T I V E   L A B E L I N G #
% # # # # # # # # # # # # # # # # # # # # # # # # # # # # # 

\subsection{P2: Visual Interactive Learning \cite{bernard2018vial}} 
\textbf{Goal:} Let human labelers visualize/explore dataset and choose instances for training accurate ML model.
\hfill\\
\textbf{Human Role:} 
Label self-selected data instances.
\hfill\\ \textbf{Problem:} 
When an ML model self-selects its own training data (as in Active Learning) there is a risk of selecting redundant information or outliers, which reduces the effectiveness of the labeling process. 
In addition, machine-centric selection does not incorporate human experience and skills. 
Much potential remains untapped in case task-educated humans are available. \hfill\\
\textbf{Structure:}
Training data is selected and annotated by the human.
Enabling human-centered selection of training instances requires a visual environment that facilitates analytical reasoning, i.e., human perception of patterns and structures in the data. 
However, manually extracting and finding associations and structures in large data sets for labeling might still be difficult.
To make hidden patterns more explicit, unsupervised ML models, in particular dimensionality reduction \cite{espadoto2019toward} can be applied. 
Dimensionality reduction is a tool for extracting low-dimensional representations of high-dimensional data while preserving the semantic relationship between all data instances. 
The low-dimensional data (usually 2 dimensions) is then visualized, e.g.~using a scatter plot.
The closer two data points are plotted, the more semantically similar they are, and vice versa.
As with Active Learning, the model is periodically retrained to update the model and visualizations. 
\hfill\\ 
\textbf{Context:} 
A user-centric data selection requiring a visual interface to facilitate the labeling process,such as a visualization tool for overlaps over documents in the dataset, outliers, tag clouds etc. Optionally, 
dimensionality reduction techniques can be used to guide the data exploration.
\hfill\\ 
\textbf{Challenges:}
\begin{itemize}
    \item Requires a visual interface to facilitate the labeling process.
    \item May requires trained humans to make sense of the visualization.
    \item Visualizations become cluttered at scale.
\end{itemize}
\textbf{Examples}:
ALTO\footnote{\url{https://github.com/Foroughp/ALTO-ACL-2016}}, 
AIDE \footnote{\url{https://github.com/microsoft/aerial_wildlife_detection}} and
mVis\footnote{\url{https://github.com/chegini91/mVis}}

% # # # # # # # # # # # # # # # # # # # # # # # # # # # # # # #
% # I N T E R A C T I V E   D A T A   A U G M E N T A T I O N #
% # # # # # # # # # # # # # # # # # # # # # # # # # # # # # # #

\subsection{P3: Trick the Model} \cite{attenberg2011beat}
\textbf{Goal:} Expose model errors to make the model more reliable against a variety of inputs. \hfill\\ 
  \textbf{Human Role:} Test the model (and trick it) with inputs that cause incorrect behavior.
  \hfill\\ \textbf{Problem:} 
Models sometimes poorly transfer learned patterns into real-world data.
Even after acquiring a reasonable number of training examples, cases may occur where a model provides unsatisfactory outcomes. 
Errors may not be in the software behavior, but rather hidden in the learned parameterization, caused by poor data quality or biased data. 
Insufficient training data causes blind spots, bias, or weaknesses against adversarial attacks to the model which has been trained. Identifying where the model might fail and closing those gaps make the model more reliable and safe to use.
\hfill\\ 
\textbf{Structure:}
Trick the Model involves two steps: a) exposing model errors and b) providing additional training instances to make the model more reliable and accurate.
%As shown in Figure 5, 
First, humans iteratively evaluate how models behave on ``manually constructed'' or slightly modified inputs (e.g. cropped imaged or edited text) to detect unintended behavior. The focus is especially on inputs, where the model is highly confident in its prediction, but is actually wrong.
If unintended behavior is found, the goal is to close the gap by collecting or creating additional training examples that address the misconception. \\
Tricking a model requires humans to understand what a model has learned and when it likely fails. 
Since it is difficult to impossible for a human to understand how a model makes decisions, an explanation mechanism that provide human-understandable insights into the behavior of the model can help (see P9), e.g.~ showing which features of the input contributed most to a class-specific decision.
\hfill\\
\textbf{Context:} 
Requires an additional post-training testing phase where model misconceptions and blind spots are identified and addressed.
 \\
\textbf{Challenges:}
\begin{itemize}
	\item Requires additional computing resources and time for training and testing.
    \item The nondeterminism in a model optimization may cause unintended side effects. 
\end{itemize}
\textbf{Examples}: Errudite\footnote{\url{https://github.com/uwdata/errudite}} and
WIT\footnote{\url{https://github.com/pair-code/what-if-tool}}

% # # # # # # # # 
% # P R O M T S #
% # # # # # # # # 

\subsection{P4: Prompt-based Learning \cite{brown2020language}}
\textbf{Goal:} Train accurate models with only a few or no task-specific training data. \\
 \textbf{Human Role:} 
Prompting the input of a pre-trained large model (PTM) to ensure either the most interesting/missing or the needed output.\hfill\\ 
 \textbf{Problem:}
While labeled data is typically the bottleneck during training, unlabeled data may also be unavailable and difficult to obtain. In this case, there is no or not enough data to learn from.
\hfill\\
\textbf{Structure:}
Prompt-based learning relies on PTMs, such as large language models (LLMs) or vision transformers, which process a general ``understanding'' of text or images. 
The first use case is to leverage the general knowledge acquired by such models to solve a specific task without additional (model) training or parameter tuning. 
The PTM is itself tuned during inference with task-specific prompts provided by humans.
In natural language processing, a prompt is a piece of text that contains a semantic description of the task.
For example, in text classification, a text \texttt{x} is wrapped into a semantic description of the classification task and passed to the model, i.e.~the input \texttt{x} is wrapped into a template such as ``\texttt{[x]} It was a \texttt{[MASK]} movie!'' 
The prediction is made based on the probability that a class-related word, i.e. ``great'' for ``sentiment positive'' or ``bad'' for ``sentiment negative'', is filled in the \texttt{[MASK]}. 
The prompt may also contain available labeled examples to tune the large model.
Another use case is to leverage PTMs as a training data
generator.
In this use-case, the PTM can be used to probabilistically fill gaps in or continue a user-provided
starting prompt, e.g. a textual phrase, according to
a desired outcome. 
For example, a LLM might be asked to fill the gap in the sentence “\textit{Overall,
it was a [\_] film}.” to satisfy a variety of sentiment levels. This outcome assist to the human labelers in the knowledge discovery. 
A secondary task-specific supervised learning model is then trained with the generated data which extends the training data.
\hfill\\
\textbf{Context:} 
Labeled and unlabeled data are not available for training a model to the necessary level. %\textcolor{red}{Uses a LLM.}
\hfill\\
\textbf{Challenges:}
\begin{itemize}
    \item Requires the use of a PLM.
    \item Could require multiple iterations of creating, refining, and analyzing prompts, which requires additional knowledge.
    \item Results could be misleading (hallucination). 
\end{itemize}
\textbf{Examples:} Visual Prompting\footnote{\url{https://app.landing.ai/public/visual_prompting}}, Prompt flow\footnote{\url{https://github.com/microsoft/promptflow}}, and PromptSource\footnote{\url{https://github.com/bigscience-workshop/promptsource}}

\section{HiL Operation Patterns}

HiL is not limited to the training-phase. It can also be effectively applied throughout the entire software lifecycle. We present four HiL operation patterns which can particularly be used when deploying and operating ML-based systems. 

% # # # # # # # # # # # # # # # # # # # # # # # # #
% # A C T I V E   D E C I S I O N   S U P P O R T #
% # # # # # # # # # # # # # # # # # # # # # # # # #

\subsection{P5: Recommendation System \cite{sengupta2017radar}} 
\textbf{Goal:}
Design the ML-based system as a recommendation system which enhances the quality and efficiency of human decisions. 
\hfill\\ 
 \textbf{Human Role:}
Make decisions based on machine recommendations.
 \hfill\\ \textbf{Problem:}
Many real-world application domains are high-stakes tasks that involve high-impact but difficult decisions, such as medical diagnosis or criminal justice. 
When the consequences of errors are significant, a purely automatic approach usually reaches its limits due to a general lack of user trust, correctness, and legal responsibility. 
In such scenarios, a domain expert (e.g. a doctor) is responsible for ``labeling'' some type of data (e.g. disease detection). 
Note that manual decision-making can also be constrained by human biases, unintentional errors, natural fuzziness and ambiguity, or can take a large amount of time. 
\hfill\\ 
\textbf{Structure:} 
To actively support a domain expert in making critical decisions, ML models can provide an initial suggestion or a list of suggestions. 
Beside a pure label, a suggestion might contain comprehensible evidence why an automated approach would select a certain outcome or how certain the prediction is. 
The support of the machine provides a ``second opinion'' and reduces the risk of humans overlooking essential things.
\hfill\\
\textbf{Context:} 
Tasks with a high degree of difficulty and large consequences where an expert has to make a decision, e.g. in health care.
\hfill\\
\textbf{Challenges:}
\begin{itemize}
	\item Requires the model to provide evidence as to why a particular recommendation might be favorable. 
	\item Risk of information (recommendation) overload or blindly trusting false evidence. 
	\item Does not scale well with large workloads, since every decision still has to be made by a human.
\end{itemize}
\textbf{Examples}: GenericCDSS\footnote{\url{https://github.com/bioinformatics-ua/GenericCDSS}},
EasierPath\footnote{\url{https://github.com/yuankaihuo/EasierPath}}% \textcolor{red}{TODO}

% # # # # # # # # # # # # # # # # # #
% # A C T I V E   I N F E R E N C E #
% # # # # # # # # # # # # # # # # # #

\subsection{P6: Active Moderation \cite{andersen2022efficient}} 

\textbf{Goal:} Improve or maintain a required level of accuracy that a model alone is unable to deliver during deployment. 
\hfill\\ 
 \textbf{Human Role:} Label a minor sequence of automatically chosen (problematic) data instances.
 \hfill\\ \textbf{Problem:} 
An already trained and fine-tuned ML model may still not be accurate enough to reach the level of accuracy required to use the model productively. Adding more training examples does not always lead to a top accuracy, as the achievable accuracy converges to a maximum as the training data increases in size. %If a higher accuracy is required, a purely manual approach would normally be needed to ensure the correctness of all predictions. 
\hfill\\  
\textbf{Structure:}
The idea of Active Moderation is to decide only those cases manually for which a model cannot provide reliable suggestions. Simple predictions, on the other hand, do not require manual verification because they are close to being correct. Active Moderation makes it possible to improve prediction performance by delegating only uncertain (and thus likely false) artificial decisions to a human oracle, while limiting human labor.
For any prediction, it is necessary to determine whether the prediction is likely to be correct (confident) or not (uncertain). If a prediction is uncertain, a human needs to give a reliable judgement instead. Since ML models generally do not make statements about their own correctness, special mechanisms are needed to quantify the reliability of the prediction. Conventionally, the prediction uncertainty is used for this purpose {\cite{andersen2022efficient}. 
\hfill\\
\textbf{Context:} 
Tasks that require a level of accuracy beyond what a machine can provide. A typical example is the moderation of online forums \cite{Loosen:SCM:2017}, i.e. deciding whether to block a user.
\hfill\\
\textbf{Challenges:}
\begin{itemize}[topsep=0pt,itemsep=0pt,parsep=0pt,partopsep=0pt]
	\item Unable to detect and avoid all model misbehavior. 
    \item Achieving near-perfect performance requires significant manual effort.
\end{itemize}
\textbf{Examples}: 
REM\footnote{\label{REM}\url{https://github.com/jsandersen/REM}},
Shipwright\footnote{\url{https://github.com/STAR-RG/shipwright}}, and Amazon Augmented AI\footnote{\url{https://aws.amazon.com/augmented-ai/}}

\iffalse
\begin{figure}
\centerline{\includegraphics[width=18.5pc]{img/pattern_am2.pdf}}
\caption{The process of active moderation. Each instance is decided by a model and the predictions' uncertainty is estimated. According to an uncertainty threshold \cite{andersen2022efficient} 
predictions are either seen as certain and  uncertain. 
A certain prediction is seen as confident and is trusted. An uncertain prediction has a high likelihood of being wrong and is thus made by a human labeler.  }\vspace*{-5pt}
\end{figure}
\fi

% # # # # # # # # # # # # # # # # # # # # # # #
% # I N T E R A C T I V E   I N V E N T I O N #
% # # # # # # # # # # # # # # # # # # # # # # #

\iffalse
\begin{figure}
\centerline{\includegraphics[width=18.5pc]{img/pattern_ii.pdf}}
\caption{
An example tool implementing the Thumps Up / Down pattern. For each ML result, human analysts can correct or agree any outcomes. 
}\vspace*{-5pt}
\end{figure}
\fi

\subsection{P7: Thumbs Up or Down \cite{shukla2021agileml}}
\textbf{Goal:} Enable fast deployments by allowing humans to correct (for them) false model outcomes when using ML-based systems. 
\hfill\\ 
 \textbf{Human Role:} 
Correct or confirm an artificial prediction while using a system.
 \hfill\\ \textbf{Problem:}
Deploying a highly accurate ML model is a lengthy process as training data is sparse and can change over time. 
Cases where ML-based predictions are incorrect can be expected, especially in early deployment phases. 
Errors identified by a human during the application of an ML system should thus be correctable. 
\hfill\\ 
\textbf{Structure:}
ML-based systems should provide the option to users to correct the outcome of the ML model, in case they are incorrect. 
The interface has to provide a mechanism to correct model outcomes. A common approach is to provide a button to overwrite, re-label, or correct any model predictions.  
\hfill\\ 
\textbf{Context:} 
ML systems for recommendations or sense-making of individual predictions.
\hfill\\
\textbf{Challenges:}
\begin{itemize}
	\item Analysts can be distracted from the actual analysis by an increased need for correction.
\end{itemize}
\textbf{Examples}:
OpenReq Analytics\footnote{\label{OpenReq Analytics}\url{https://github.com/OpenReqEU/ri-visualization}},
DeepL Write\footnote{\url{https://www.deepl.com/de/write}}, and 
Google Translate\footnote{\url{https://translate.google.com/}}

% # # # # # # # # # # # # # # # # # # # # # # #
% # I N T E R A C T I V E   I N V E N T I O N #
% # # # # # # # # # # # # # # # # # # # # # # #

\subsection{P8: Continuous Learning \cite{karmakharm2019journalist}} 
\textbf{Goal:} Regularly update the ML model to ensure it continues to provide accurate and reliable predictions. 
\hfill\\ 
 \textbf{Human Role:} 
 Continuously label new data instances.
 \hfill\\ \textbf{Problem:} 
Real-world data is highly dynamic and prone to changes. 
For ML models, it is critical that the training data matches the current state of the environment in which a system is being used. 
If a model is applied to new data that is significantly different from the training data, the model will struggle to recognize patterns that it has not seen before, leading to incorrect results. 
\hfill\\ 
\textbf{Structure:}
Deployed ML models need to be updated with  data that reflect its current environment to maintain their effectiveness. Continuous Learning aims to retrain a model on a regular basis when additional training data is available, e.g.~daily or when a certain amount of additional labeled data is available. 
By doing so, the model learns new patterns and trends and continuously adapts to changes in the environment, i.e. new terms, slang, image resolutions, laws, etc. 
Training data 
is extended with labeled examples that arise during the use of the systems. 
\hfill\\
\textbf{Context:} 
ML systems in dynamic environments.
\hfill\\
\textbf{Examples}: Forum 4.0\footnote{\url{https://git.informatik.uni-hamburg.de/mast/showcase/projects/forum40/forum40}}, 
OpenReq Analytics\footref{OpenReq Analytics}, and REM\footref{REM}

\section{HiL Collaboration Pattern}

HiL patterns for ML-based systems rely heavily on the effective cooperation between humans and ML models through the exchange of labels and feedback. 
Collaboration Patterns describe proven ways in which the generic exchange between human feedback and machine prediction can be supported.  
We outline two additional orthogonal HiL patterns to enable a focused and effective humans machine collaboration.

\subsection{P9: Instance-based Explanation \cite{bhatt2020explainable}}
\textbf{Goal:} Maintain user trust in ML models by providing evidence and insight as to why and how a particular decision was made.
\hfill\\ 
 \textbf{Problem:} 
Most ML models are input-output machines that are considerable a “black box”, as they do not reveal human comprehensible insights into their decision-making process. However, users (particularly domain experts) are usually unwilling to rely on artificial predictions from ML models if they cannot trust them, especially in high-stakes tasks. Furthermore, data labeling is a time-consuming and tedious process and not exploiting the knowledge accumulated by models is a missed opportunity for HiL applications. 
\hfill\\ 
\textbf{Structure:}
Communicating how models arrive at each individual decision is a desired property when relying on ML outcomes. 
An instance-based explanation of a model prediction is any kind of human-oriented information that provides confidence in an individual model outcome. In general, there are two generic types of instance-based explanations: relevance scores and uncertainties,  which can be extracted from common classifiers with specific calculation  tools \cite{andersen2022efficient,bhatt2020explainable}. Relevance scores highlight which features of the input contribute most to a particular outcome. These indicate which features the model has considered the most when making predictions, i.e.~pixels of an image or worlds of a text. 
Relevance scores indicate why a certain decision was made. 

Uncertainty refers to the variability and undecidability of a prediction. The higher the uncertainty, the more difficult it was for the model to make a clear decision. Uncertainty scores indicate how reliable a prediction is. 
Relevance scores and uncertainties can be presented to human labelers to affect the labeling process, increase user trust and to make explicit what a model knows and not knows. 
\hfill\\
\textbf{Context:} 
ML-based systems where user trust is desired or needed. 
Explanations can be combined with the moderation, tricking or the recommendation pattern.
\hfill\\
\textbf{Examples}: LIME\footnote{\url{https://github.com/marcotcr/lime}},
SHAP\footnote{\url{https://github.com/shap/shap}}, and 
keras-uncertainty\footnote{\url{https://github.com/mvaldenegro/keras-uncertainty}}

\subsection{P10: Crowd Agreement \cite{plank2022problem}} 
\textbf{Goal:} To diversify perspectives and increase the accuracy and consistency of human labels. 
 \hfill\\ \textbf{Problem:}  
Although the amount of human labeled data is typically small and might include noise and biases, it is used for the crucial task of extrapolating patterns to unseen data. 
Also, humans can be biased and inaccurate. Their labels are expected to vary according to their level of attention, knowledge, and experience \cite{plank2022problem}. 
\hfill\\ 
\textbf{Structure:}
Asking multiple humans for their opinions is a general approach to reducing the variance in the error of human-provided labels. 
Crowd agreements aim at involving several independent humans in a HiL workflow and then resolving conflicts according to an agreement strategy. An agreement measures the degree to which multiple labelers make the same decision. It is used to identify the best overall decision. 
Advanced agreements strategies typically take into account human experience and skill levels, which have been shown to have a significant impact on label quality. 
\hfill\\
\textbf{Context:} 
ML systems in which several labelers are used to label the same instances in order to increase reliability. 
\hfill\\
\textbf{Examples}: UIMA-Agreement\footnote{\url{https://github.com/texttechnologylab/UIMA-Agreement}},
Cohen Kappa\footnote{\url{https://scikit-learn.org/stable/modules/generated/sklearn.metrics.cohen_kappa_score.html}}, and 
bratiaa\footnote{\url{https://github.com/kldtz/bratiaa}}

\section{DISCUSSION and CONCLUSION}
Deployed ML systems are to some extent imperfect and lack reliability and accuracy. 
They may not be able to solve difficult tasks. 
Looping the feedback of users and domain experts into the development and deployment processes emerges to overcome these limitations.
We propose a catalog of HiL training, operation, and collaboration patterns describing best practices and proven solutions.

Patterns for accumulating the initial training data (P1, P2, P3, and P4) aim at reducing the number of labels needed (and human involvement) to train accurate models. 
Frequent retraining of models can be very costly and time-consuming (P8). 
As labeling a sequence of machine-selected instances (P1) is often perceived as exhausting, visual approaches (P2) support labelers  gain additional knowledge about the data through data visualization. 
However, users must be able to interpret these visualizations (P2, P3, P4) which might require additional skills, e.g., to understand semantic relationships or similarity of inputs.

One central problem is that predictions might be considered trustworthy when they are not. While adequate explanations can build trust (P10), there is also a danger of convincing humans of a false correctness. 
Putting a human into the operational loop of an ML-based system has been shown to improve its accuracy and reliability (P5, P6 and P7). 
However, the question is whether the human resources are available to the extent required and whether the added value is worthwhile, i.e. whether the HiL approach significantly improves reliability compared to automatic predictions. 
One direction it to explicitly capture human feedback directly when working with a system in the form of corrections (P7). 
It should be noted, however, that human labelers, like machines, are prone to errors and may provide incorrect labels (P9).  

The described patterns can and certainly should be combined.
The proposed catalog is not meant to be complete. It is the results of our own experience and a non-systematic literature review. Further work should extend the catalog and evaluate its suitability and utility in practice.

\def\refname{REFERENCES}

\bibliographystyle{IEEEtran}
\bibliography{hil-patterns}

\end{document}